# Intrinsic ultrafast edge photocurrent dynamics in $WTe_2$ driven by broken crystal symmetry

Subhashri Chatterjee,[1] Katsumasa Yoshioka,[1,*] Taro Wakamura,[1]
Vasili Perebeinos,[2] and Norio Kumada[1]

[1]Basic Research Laboratories, NTT, Inc., 3-1 Morinosato-Wakamiya, Atsugi, 243-0198, Japan
[2]Department of Electrical Engineering, University at Buffalo, Buffalo, New York, 14260, United States
*e-mail: katsumasa.yoshioka@ntt.com

**Abstract: Directional photocurrents in two-dimensional materials arise from broken crystal symmetry, offering pathways to high-speed, bias-free photodetection beyond conventional devices. Tungsten ditelluride ($WTe_2$), a type-II Weyl semimetal, exhibits robust symmetry-breaking-induced edge photocurrents from competing nonlinear optical and photothermoelectric mechanisms, whose intrinsic dynamics have remained experimentally inaccessible. Here, we directly resolve sub-picosecond edge photocurrent dynamics in $WTe_2$ through ohmic contacts over temperatures from 300 K to 4 K. We demonstrate ultrafast optical-to-electrical conversion with a 3 dB bandwidth of ~250 GHz and reveal picosecond-timescale switching of the net photocurrent direction below 150 K, linked to a Lifshitz transition. This transient bipolar response arises from non-equilibrium Seebeck effects due to asymmetric cooling of hot electrons and holes. These findings reveal previously hidden ultrafast dynamics in symmetry-engineered materials, offering new strategies to disentangle competing photocurrent mechanisms and enabling the development of self-powered, ultrafast optoelectronic devices.**

**Main text:**

Breaking crystal inversion or rotational symmetry in two-dimensional (2D) materials offers a powerful route to generate directional photocurrents,[1-18] without the need for conventional p–n junctions or external bias. These photocurrents are selectively sensitive to wavelength and polarization with tunable current direction, which not only enable novel optoelectronic functions, including on-chip polarimetry[3] and broadband neuromodulation,[18] but also hold promise for optical-to-electrical (O–E) conversion efficiencies beyond the Shockley–Queisser limit.[2,8-9,11-13,19] However, symmetry breaking can activate multiple competing photocurrent mechanisms, complicate their understanding and limit practical applications.

Tungsten ditelluride ($WTe_2$), a type-II Weyl semimetal[20-22] with a large Berry curvature and low crystal symmetry, offers an ideal platform for exploring such effects.[5-7,23-25] At edges misaligned with the *a*- and *b*-axes, where mirror symmetry is broken, directional photocurrents emerge without external fields. Scanning photocurrent microscopy (SPCM) studies suggest a nonlinear optical origin[5,7] such as the shift current, potentially enhanced by topological Fermi-arc states.[5] In contrast, photocurrent-flow microscopy, which visualize local photocurrent using high-sensitive quantum magnetometry with nitrogen-vacancy canter spins, indicates a photothermoelectric (PTE) effect arising from anisotropic Seebeck coefficients along the crystal axes ($S_a \neq S_b$).[6] These mechanisms can be disentangled by their distinct intrinsic dynamics: shift currents respond instantaneously to optical excitation,[26-31] while PTE currents evolve over picosecond timescales due to hot carrier cooling.[32-35] Although terahertz (THz) emission spectroscopy[26–29,31,36–39] can probe local ultrafast polarization dynamics, it cannot reveal how such microscopic processes translate into net charge flow delivered to an external circuit. Direct electrical readout via ohmic contacts is therefore indispensable for capturing the actual photocurrent waveform and unambiguously determining both the speed and mechanism of the O–E conversion process.

Beyond symmetry, the underlying electronic structure also governs net photocurrent flow. Notably, $WTe_2$ exhibits a temperature-dependent Lifshitz transition that reshapes its Fermi surface, creating an electron-hole compensated regime below ~150 K[40–42] where unconventional[43–45] and topological[46] transport phenomena emerge. Tracking the temperature dependence of intrinsic photocurrent dynamics sheds light on how symmetry and electronic topology cooperatively dictate ultrafast carrier transport, yet this fundamental connection has remained experimentally unexplored.

Here, we directly resolve sub-picosecond edge photocurrent dynamics in $WTe_2$ using a laser-triggered photoconductive (PC) switch[32,47–49] over 4–300 K. Ultrafast O–E conversion with a ~250 GHz 3 dB bandwidth emerges exclusively at mirror-symmetry-breaking edges, the fastest reported in Weyl semimetals. Below the Lifshitz transition (<150 K), the photocurrent waveform evolves from a single peak to a bipolar shape, indicating picosecond-timescale switching of the net current direction. This transient bipolar photocurrent arises from a non-equilibrium Seebeck effect due to asymmetric cooling of hot electrons and holes. The picosecond decay, its strong temperature dependence, and systematic trends with polarization, power, and wavelength confirm a PTE origin over nonlinear optical mechanisms. These findings offer the first direct view of ultrafast edge photocurrent dynamics in a Weyl semimetal and present a broadly applicable framework for symmetry-engineered optoelectronics.

## Experimental setup

Figure **1A** illustrates our experimental setup.[32] $WTe_2$ was mechanically exfoliated and then a 150-nm-thick flake was transferred onto a sapphire substrate. Source and drain titanium/gold (Ti/Au) electrodes were deposited on top of the flake. The drain electrode forms a Goubau-line waveguide, connecting the $WTe_2$ flake to a low-temperature grown gallium arsenide (LT-GaAs) PC switch placed 440 μm away. To demonstrate the reproducibility, we fabricated another sample (sample B) and included the results in Supporting Information (Figure **S2**). A femtosecond pulsed laser with a pulse duration of 280 fs was employed to excite and detect the photocurrent. The pump pulse was tightly focused on $WTe_2$, while a probe pulse was focused on the PC switch with a controlled delay $t$ to measure the photocurrent in time domain. The position of the pump beam was controlled using a motorized mirror. All measurements except for the wavelength dependence were performed using 517 nm light. The sample was placed in a cryostat with a base temperature of 4 K. Throughout the experiments, we kept the source-drain bias zero to focus on the photocurrent originating from broken crystal symmetry.

The inset of Figure **1A** displays the orthorhombic crystal structure of $T_d$-$WTe_2$, which belongs to the space group $Pmn2_1$. Zigzag chains of tungsten atoms along the $a$-axis lead to anisotropic electrical and thermal transport.[50–52] Here, mirror symmetry is broken at edges not aligned with the principal crystallographic directions. The crystallographic axes, determined by Raman spectroscopy (Figure **S1**), are indicated in Figures **1B** and **C**.

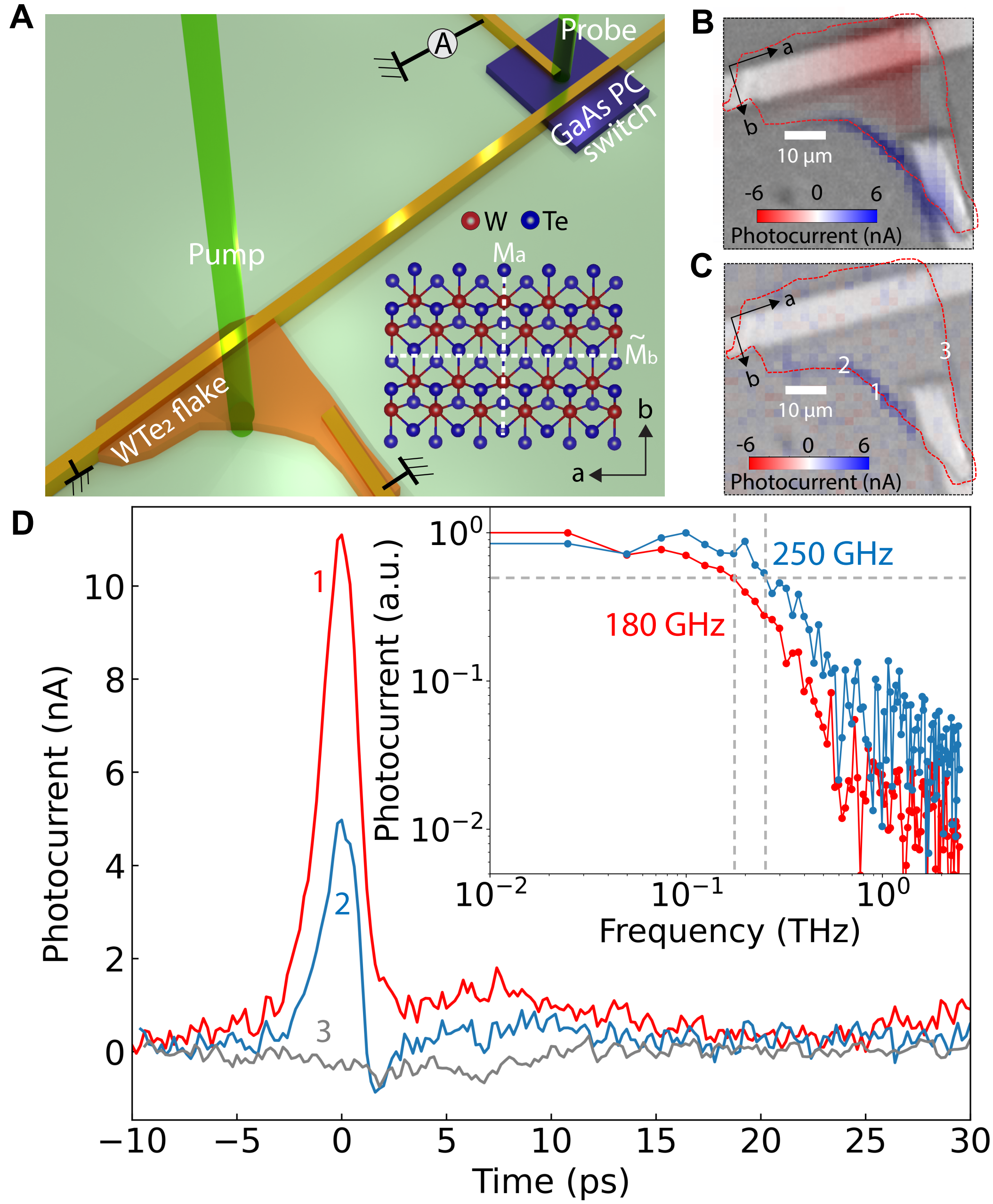

**Figure 1 | On-chip ultrafast edge photocurrent measurement.** **A**, Schematic of the $WTe_2$ based on-chip photodetector device, where the LT-GaAs photoconductive switch is connected via the Goubau-line waveguide. Inset shows the crystal structure of $T_d$-$WTe_2$. $M_a$ and $\widetilde{M_b}$ represent the mirror and glide-mirror planes, respectively. **B**, Time-integrated photocurrent image at 300 K. **C**, Transient photocurrent image at zero-time delay ($t$ = 0) at 300 K. **D**, Photocurrent as a function of $t$ at several pump positions indicated by numbers in (**C**). Inset shows normalized Fourier-transformed spectra at positions 1 and 2. 3 dB bandwidth are shown by vertical dashed lines: 180 and 250 GHz for positions 1 and 2, respectively.

## Experimental results

### Time-domain measurement of edge photocurrent

Figure **1B** presents the time-integrated photocurrent map measured at 300 K. Consistent with the previous continuous-wave-laser-based SPCM measurement,[5–7] photocurrent is generated not only when the pump beam is focused on the electrodes, but also when it is focused on the mirror-symmetry-breaking edges, which are noticeably misaligned with the *a*- and *b*-axes. These two contributions can be disentangled by time-resolved photocurrent measurement. Figure **1C** presents a transient photocurrent map at zero-time delay ($t = 0$) at a laser fluence of 0.68 mJ/cm$^2$. An instantaneous positive photocurrent is observed exclusively along the mirror-symmetry-breaking edges. This observation reveals two key insights that are accessible only through direct readout of ultrafast photocurrent. First, the edge photocurrent is generated under femtosecond-pulse excitation within a sub-picosecond timescale, indicating an ultrafast response. Second, it originates from a mechanism fundamentally distinct from that responsible for the photocurrent near electrodes. The slow response and the opposite current directions observed near the two electrodes can be attributed to photovoltaic effects associated with the built-in electric fields at the $WTe_2$-electrode interfaces.[53]

Figure **1D** compares the photocurrent as a function of *t* at different excitation spots under a laser fluence of 0.68 mJ/cm$^2$. The full width at half maximum (FWHM) of the edge photocurrent is approximately 1 to 1.8 ps, highlighting the ultrafast carrier dynamics in $WTe_2$. A Fourier transform analysis yields a 3 dB bandwidth of 180 and 250 GHz at positions 1 and 2, respectively (inset of Figure **1D**). These values represent the fastest electrical response reported for $WTe_2$-based photodetectors[54–57] and are competitive with those other state-of-the-art 2D material-based photodetectors.[32,58–59] This result highlights the potential of $WTe_2$ as an ultrafast O-E conversion platform. For the subsequent measurements, we fixed the excitation spot at position 1, which exhibits the highest photocurrent amplitude.

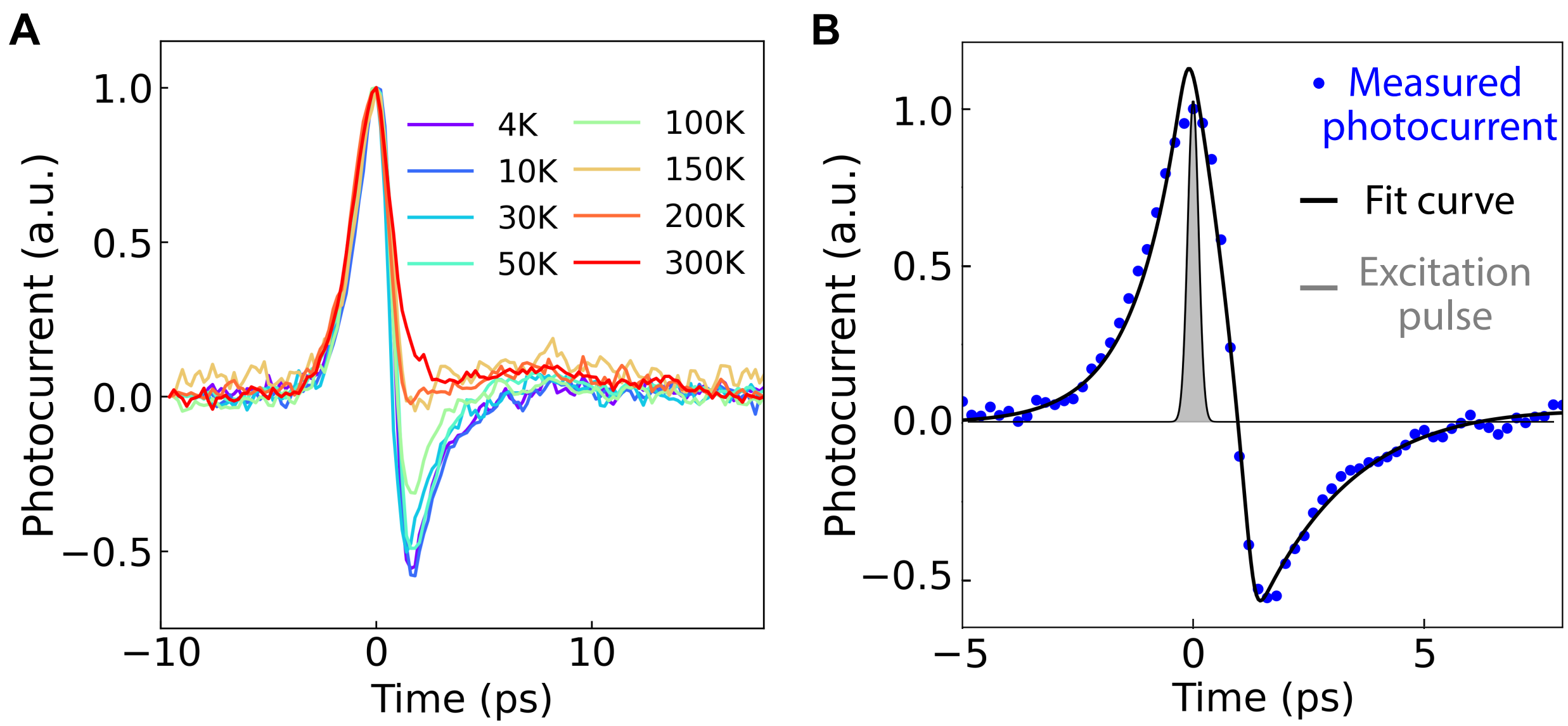

**Figure 2 | Temperature variation of edge photocurrent waveform. A,** Photocurrent at position 1 for various temperatures under a laser fluence of 0.68 mJ/cm$^2$. Traces are normalized to the maximum current. **B**, Fitted time domain data at 4K overlapped with a gaussian 280 fs excitation pulse.

### Temperature dependent photocurrent waveform

To elucidate the origin of the edge photocurrent, we investigate its temperature dependence. Figure **2A** shows the time-resolved photocurrent traces measured at position 1 over a temperature range from 300 K to 4 K. The photocurrent at 300 K exhibits a single positive peak, indicating unidirectional current transport. As the temperature decreases, a negative peak begins to emerge around 100 K following the initial positive peak, indicating an ultrafast reversal of current direction. Upon further cooling, the relative amplitude of the negative peak increases and becomes nearly temperature-independent below 50 K.

Figure **2B** shows the waveform analysis to better understand the origin of this ultrafast photocurrent. The 4 K time-domain waveform was fitted by accounting for the convolution effects of the laser pulse duration and detection response of the PC switch, with the photocurrent decay times of the positive and negative peaks used as fitting parameters (see Supporting Information for details, Section I).[47] The decay times of the positive and negative components were estimated to be 2.8 ± 0.1 ps and 3.8 ± 0.2 ps, respectively, more than an order of magnitude longer than the 280-fs excitation pulse duration. This observation, combined with the pronounced temperature dependence of the waveform, rules out a shift-current origin. This mechanism arises from coherent interband excitation that induces a real-space displacement of the electron wave packet characterized by the shift vector. Consequently, the photocurrent follows the pulse intensity envelope and typically vanishes within femtosecond timescales due to dephasing and scattering.[26–31] In contrast, the observed time scale is consistent with the PTE effect, where the photocurrent decay time is determined by the cooling time of photoexcited carriers,[32–35] governed by electron-phonon interactions with a characteristic timescale on the order of picosecond in $WTe_2$.[60–62] Therefore, the observed edge photocurrent is due to an anisotropic Seebeck coefficient. In this case, the photocurrent amplitude is expected to be proportional to $S_a - S_b$ where $S_a$ and $S_b$ are Seebeck coefficients along the *a*- and *b*- axes, respectively.[6] Additional evidence supporting the PTE origin will be discussed later in conjunction with Figure **4**.

Having identified the PTE origin of the edge photocurrent, we discussed the mechanism underlying the transient bipolar photocurrent in relation to the electronic structure of $WTe_2$. The observed pronounced temperature dependence can be attributed to the change in the carrier dynamics associated with the temperature-induced Lifshitz transition.[40-42] In this case, the Lifshitz transition refers to the emergence of hole pockets in addition to electron pockets below ~150 K as the Fermi energy, which is located above the valence-band maximum at room temperature, shifts downward with decreasing temperature.[40–42] This transition temperature matches the temperature at which the negative peak in the photocurrent waveform begins to emerge. This suggests that the development of the bipolar structure at low temperatures is linked to the coexistence of electron and hole pockets. In general, since electrons and holes contribute oppositely to the Seebeck effect, the sign of $S_a - S_b$ and, consequently, the photocurrent polarity can vary dynamically depending on their relative populations and relaxation times.

To clarify this behavior, we developed a theoretical model for the Seebeck coefficient in a semimetal, incorporating the distinct carrier temperatures in the electron and hole pockets. Based on density functional theory (see Supporting Information for details, Section II), we calculated the non-equilibrium Seebeck coefficients $S_{a(b)}$ along the *a*- and *b*-axes of bulk $WTe_2$. $S_{a(b)}$ was derived using the following expression:

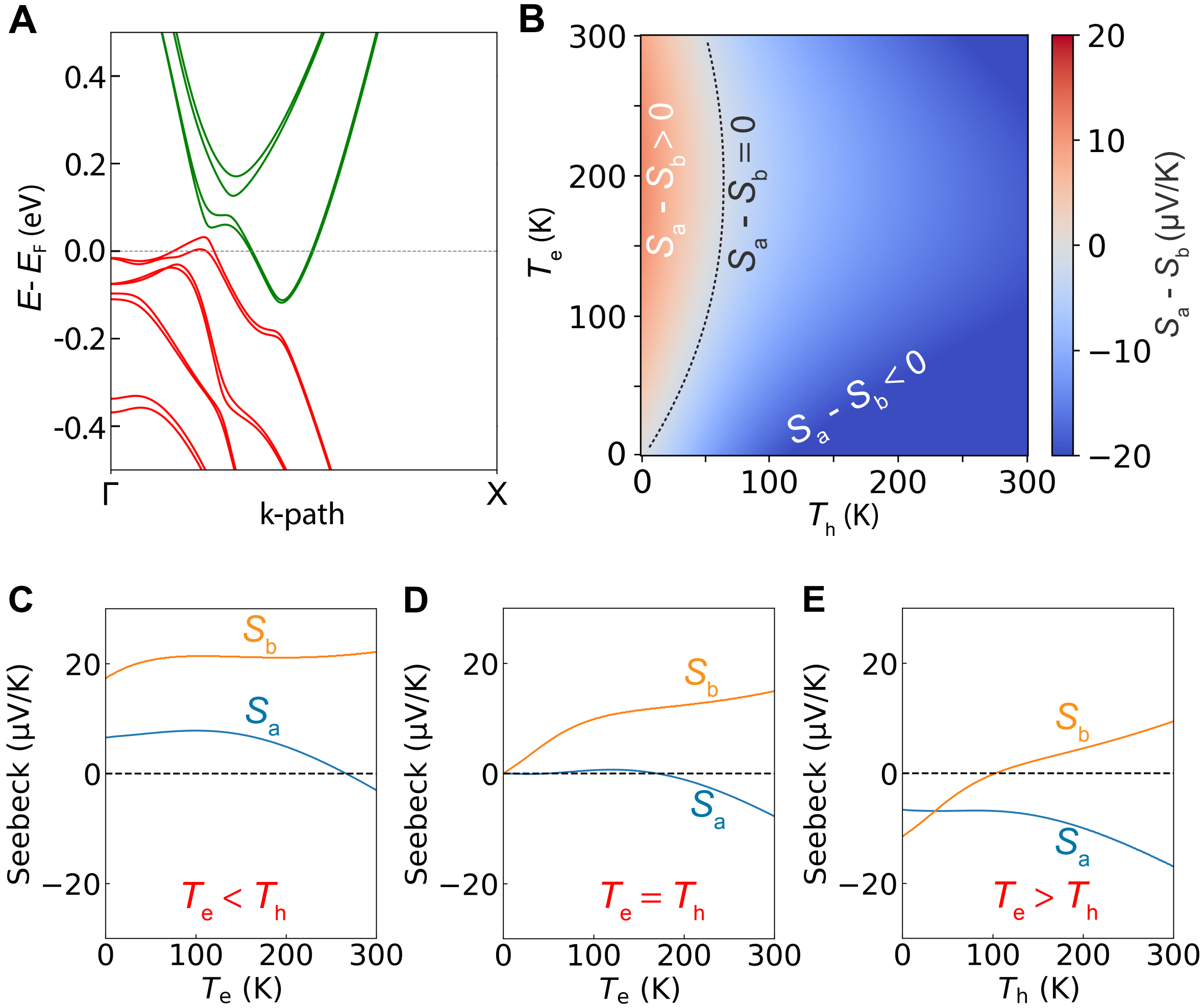

**Figure 3 | Calculation of non-equilibrium Seebeck coefficient. A**, Calculated band structure of $WTe_2$. Dashed line represents the Fermi energy ($E_F$) assumed for the Seebeck coefficient calculations. **B**, 2D contour plot of $S_a$ - $S_b$ as a function of carrier temperatures in the electron ($T_e$) and hole ($T_h$) pockets. Calculated Seebeck coefficients $S_a$ and $S_b$ for $T_h = T_e + 50$ K (**C**), $T_e = T_h$ (**D**), and $T_e = T_h + 50$ K (**E**).

$$S_{a(b)} = -\frac{1}{e}\frac{\int_{E_c}^{\infty}\left(\frac{E-\mu}{T_e}\right)\left(-\frac{\partial f_e}{\partial E}\right)\sigma_{ea(b)}(E)dE+\int_{-\infty}^{E_v}\left(\frac{E-\mu}{T_h}\right)\left(-\frac{\partial f_h}{\partial E}\right)\sigma_{ha(b)}(E)dE}{\int_{E_c}^{\infty}\left(-\frac{\partial f_e}{\partial E}\right)\sigma_{ea(b)}(E)dE+\int_{-\infty}^{E_v}\left(-\frac{\partial f_h}{\partial E}\right)\sigma_{ha(b)}(E)dE} \quad (1)$$

Here, $e$ is the elementary charge, $\mu$ is the chemical potential, $f_e$ and $f_h$ are the Fermi-Dirac distribution functions for electrons and holes, respectively. The electron and hole temperatures are denoted by $T_e$ and $T_h$ respectively, allowing for the inclusion of non-equilibrium carrier dynamics.

The energy-dependent conductivities of electron $\sigma_{ea(b)}(E)$ and hole $\sigma_{ha(b)}(E)$ are resolved along the respective crystallographic directions and are defined by the following expression:

$$\sigma_{a(b)}(E) = \sum \int_{BZ} \tau_n(\vec{K}) v_{na(b)}^2(\vec{K}) \delta(E - E_n(\vec{K})) d\vec{K}, \quad (2)$$

where $\tau_n(\vec{K})$ represents the momentum-dependent relaxation time, $v_{na(b)}(\vec{K})$ is the group velocity component along the *a*- or *b*-axis, $E_n(\overrightarrow{K})$ denotes the energy dispersion for band *n*, and the integral spans the Brillouin zone (BZ). As shown in Figure **3A**, we set the Fermi level at the charge neutrality point (dashed line), thereby highlighting the electron-hole compensated regime below the Lifshitz transition. The valence bands (red) and conduction bands (green) are emphasized to illustrate the coexistence of hole and electron contributions.[40] Figure **3B** shows $S_a - S_b$ as a function of $T_e$ and $T_h$. In thermal equilibrium ($T_e = T_h$), $S_a - S_b$ remains negative over the entire temperature range, whereas under non-equilibrium conditions with $T_e > T_h$, a region emerges where the sign of $S_a - S_b$ becomes positive. The dotted curve marks the $S_a - S_b = 0$ boundary. This behavior is further clarified by the plot of $S_a$ and $S_b$ under three representative carrier temperature conditions: $T_e < T_h$, $T_e = T_h$, and $T_e > T_h$ (Figures **3C–E**). When $T_e = T_h$, electron and hole contributions nearly cancel along the *a*-axis ($S_a \approx 0$), while holes dominate along the *b*-axis ($S_b > 0$), leading to $S_a - S_b < 0$. In contrast, when $T_e > T_h$, enhanced electron contributions shift both $S_a$ and $S_b$ toward negative values, with a larger shift in $S_b$. This results in a reversal in the sign of $S_a - S_b$ at low temperatures. Conversely, when $T_e < T_h$, the stronger hole contribution significantly increases $S_b$, and $S_a - S_b$ remains negative.

These calculations indicate that non-equilibrium carrier heating and cooling in the electron and hole pockets can cause a dynamic reversal of $S_a - S_b$, and consequently, the photocurrent polarity. This provides an explanation for the bipolar waveform observed below 150 K, in which a negative peak follows a positive one: immediately after the photoexcitation, the system resides in the $S_a - S_b < 0$ region, but as cooling proceeds, it crosses into the $S_a - S_b > 0$ region. This is plausible because the specific heat and carrier-phonon coupling may differ between the electron and hole pockets, leading to different $T_e$ and $T_h$ immediately after thermalization as well as distinct cooling rates for each. Notably, such transient anisotropy in the Seebeck coefficient arises only under non-equilibrium conditions and remains inaccessible to conventional steady-state or time-integrated measurements. Our results, therefore, demonstrate that ultrafast, time-resolved photocurrent readout is essential for revealing these hidden dynamics. By disentangling the temporal evolution of electron and hole temperatures along distinct crystallographic directions, we establish a robust framework for probing non-equilibrium thermoelectric phenomena in materials with broken crystal symmetry.

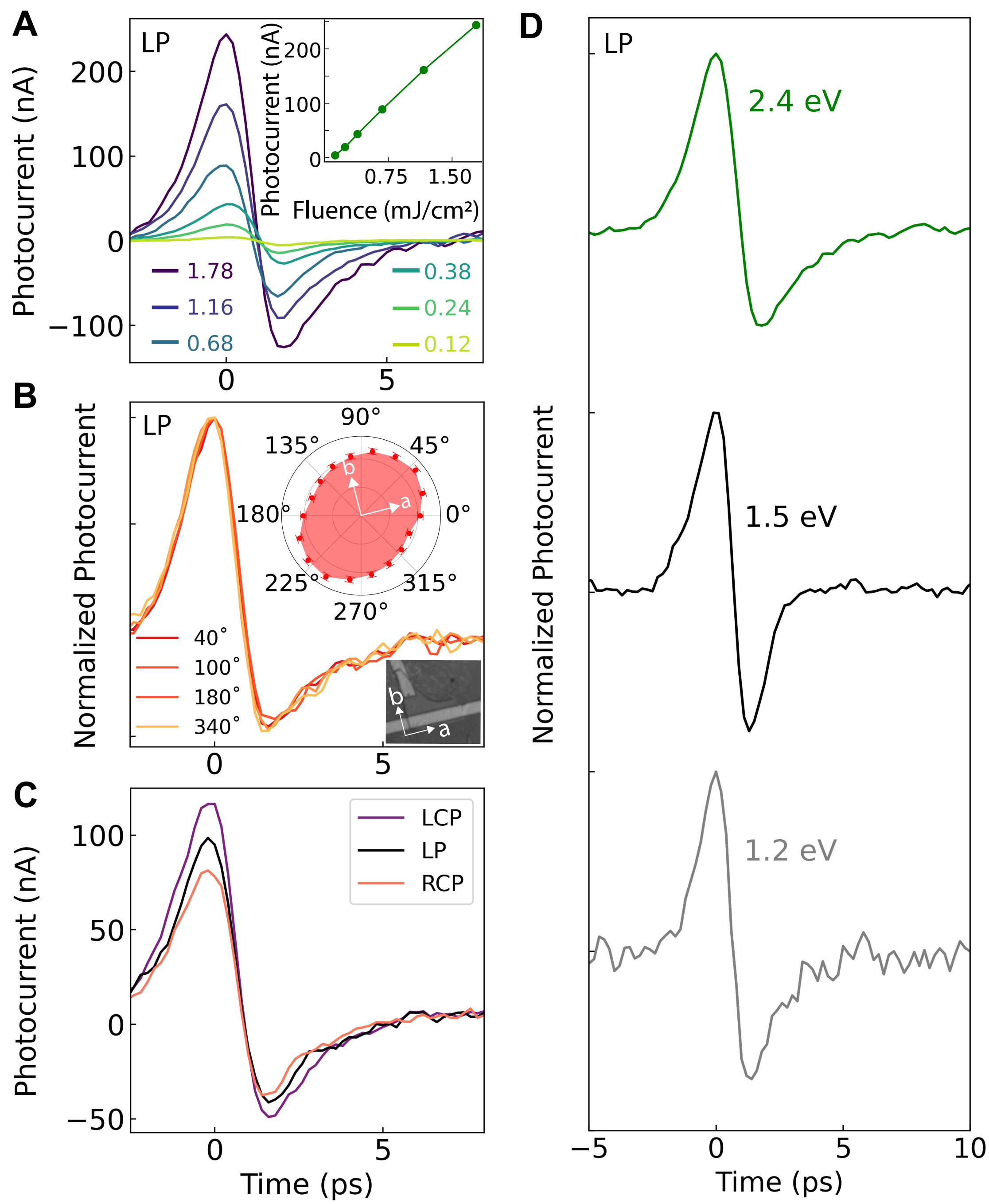

**Figure 4 | Dependence of time-domain photocurrent response on light intensity, polarization angle, and wavelength at 4 K**. **A**, Dependence on light intensity (inset: extracted positive peak amplitude). **B**, Polarization angle dependence under linearly polarized (LP) light (inset: polar plot of the positive peak amplitude). **C**, Response under left and right circularly polarized (LCP and RCP) light excitation. **D**, Wavelength dependence for 517 nm (0.68 mJ/cm$^2$), 780 nm (0.25 mJ/cm$^2$) and 1035 nm (1.01 mJ/cm$^2$). Each waveform is offset by two for clarity.

To further confirm the PTE origin of the edge photocurrent, we examined its dependence on light intensity, polarization, and wavelength at 4 K (corresponding 300 K data are provided in Figure **S3-6** in the Supporting Information). As shown in Figure **4A**, the photocurrent increases linearly with incident fluence, consistent with the PTE effect[34], and shows no saturation, one of the signatures of the shift-current mechanism.[1–2,16,26–27] Figure **4B** shows the dependence on polarization angle of linearly polarized (LP) light. Although the amplitude modulates slightly (Figure **S5**), the waveform and current direction remain unchanged, in contrast to the reversal expected from carrier-momentum alignment, another nonlinear optical process.[7,63] Under left and right circularly polarized (LCP and RCP) light (Figure **4C**), the photocurrent amplitude shows only minor modulation due to spin contributions,[64] but the waveform remains unchanged, further ruling out a shift current origin. The waveform was likewise unaffected by varying the excitation wavelength (517, 780, and 1035 nm; photon energies of 2.4, 1.5, and 1.2 eV) (Figure **4D**). All these observations are consistent with PTE dynamics governed by hot-carrier cooling, which are insensitive to both polarization and incident photon energy.

## Discussion

In this work, we performed on-chip ultrafast photocurrent readout from the symmetry-breaking edge of $WTe_2$ across a temperature range from 300 K down to 4 K. The observed photocurrent arises from the material's intrinsic low symmetry, without the application of any external bias. Based on its picosecond-scale temporal response, pronounced temperature-dependent waveform, linear amplitude scaling with pump intensity, and insensitivity to both polarization and excitation wavelength, we conclude that the edge photocurrent in $WTe_2$ originates from an anisotropic PTE effect.

We demonstrate that O-E conversion in $WTe_2$ exhibits a 3 dB bandwidth of 250 GHz at room temperature, representing the fastest performance recorded in Weyl semimetals. By combining sub-picosecond time-resolved photocurrent measurements with theoretical modeling of non-equilibrium anisotropic Seebeck coefficients, we uncover a mechanism in which a transient imbalance between electron and hole temperatures leads to ultrafast reversal of photocurrent flow at low temperatures. This behavior is directly linked to the temperature-induced Lifshitz transition below ~150 K, where electron and hole pockets coexist, highlighting the inherently temperature-tunable nature of $WTe_2$-based optoelectronics. These findings emphasize the critical role of symmetry breaking and the unique non-equilibrium carrier dynamics associated with the semi metallic nature of $WTe_2$ in governing its ultrafast photocurrent transport.

Our approach of resolving intrinsic photocurrent dynamics not only reveals fundamental aspects of photocurrent generation at symmetry-broken edges but also establishes a versatile platform for clarifying other symmetry-driven photocurrent phenomena in 2D materials, such as those arising in van der Waals heterostructures,[1-2] moiré superlattices,[3-4] uniaxial strained systems[9–11] and grain-boundaries.[12] Detailed insights into the underlying photocurrent generation mechanisms and their temporal dynamics provide critical guidelines for designing practical optoelectronic devices with optimized bandwidth and efficiency, crucial for surpassing the capabilities of conventional semiconductor-based technologies.

## Methods

### Device fabrication

Bulk $WTe_2$ crystals were mechanically exfoliated and transferred directly onto 10 × 10 mm sapphire substrates. A 2.6-μm-thick LT-GaAs chip, which was prepared using a wafer supplied by BATOP GmbH, was then transferred onto the sapphire substrate using a thermoplastic methacrylate copolymer (Elvacite 2552C, Lucite International) as an adhesive.[65] Residual adhesive was removed by rinsing the substrate in chloroform for 1 hour, followed by acetone for 1 minute and isopropyl alcohol (IPA) for 1 minute. Ti/Au waveguides and contacts were deposited using vacuum evaporation. All exfoliation, transfer, and cleaning processes were conducted in a clean-room environment under ambient conditions.

### On-chip time-domain photocurrent measurements

Femtosecond laser pulses at 1035 nm (Monaco, Coherent Ltd.), 780 nm (C-Fiber 780, Menlo Systems GmbH), and 517 nm (generated via second harmonic generation using a beta-barium borate crystal) were used to excite the device. We selected 517 nm for better focusing, achieving a spot size of 1.8 μm (FWHM), compared to 2.4 μm and 3.6 μm for 780 nm and 1035 nm, respectively. The repetition rates were 16.8 MHz for 517 nm and 1035 nm, and 150 MHz for 780 nm. Pump and probe beams were combined using a beam splitter and carefully aligned with a slight spatial offset. This allowed them to be focused onto the $WTe_2$ and the LT-GaAs PC switch, respectively, through an objective lens. The pump beam was horizontally polarized except for polarization-dependent measurements. The optical power and polarization angle were controlled using a half-wave plate. Its position was adjusted using a motorized mirror system, whereas the probe beam remained fixed throughout the experiments. To enable lock-in detection of the THz signal, the pump beam was modulated using an optical chopper operating at several hundred Hz. All experiments were conducted under a vacuum environment to minimize environmental influences and prevent degradation of the sample. The pump fluence reported in this study refers to the peak fluence of a Gaussian beam, defined as $F_0 = 2E_{pulse}/(\pi w^2)$, where $E_{pulse}$ is the pulse energy and $w$ is the $1/e^2$ radius.

### DC photocurrent SPCM measurements

Time-integrated SPCM measurements were performed using femtosecond lasers at 517 nm, 780 nm, and 1035 nm, as well as a continuous-wave (CW) laser at 780 nm. The laser beam was focused on to the $WTe_2$ flake and scanned in the x and y direction for recording spatial distribution of the DC photocurrent. Photocurrent was measured by a lock-in amplifier while another electrode was grounded. Both SPCM and time-domain data was taken under zero source-drain bias conditions.

### Acknowledgement

We thank H. Murofushi for technical support.

### Funding Sources

This work was supported by JSPS KAKENHI Grant number JP24H00828. This work was also conducted as a part of ELEQUANT project, which has received funding from European Innovation Council and SMEs Executive Agency under Grant Agreement 101185712. V.P. acknowledges US-JAPAN PIRE collaboration, National Science Foundation Grant No. 2230727, and the JSPS short-term visit program Grant No. S24109.

### Author contributions

K.Y. and N.K. conceived the experiment. K.Y. designed and built the experimental set-up. S.C. performed the measurement. S.C. K.Y. analyzed the data. S.C. K.Y. and N.K. designed the THz circuits. S.C. and T.W. fabricated the devices. V.P. calculated the Seebeck coefficient based on density functional theory. S.C. K.Y. and N.K. wrote the paper, with input from all authors.

### Competing interests

The authors declare no competing interests.

### Data availability

The datasets generated during and/or analyzed during the current study are available from the corresponding author on reasonable request.

**Supplementary Information for**

# Intrinsic ultrafast edge photocurrent dynamics in $WTe_2$ driven by broken crystal symmetry

Subhashri Chatterjee,[1] Katsumasa Yoshioka,[1,*] Taro Wakamura,[1]
Vasili Perebeinos,[2] and Norio Kumada[1]

[1]Basic Research Laboratories, NTT, Inc., 3-1 Morinosato-Wakamiya, Atsugi, 243-0198, Japan
[2]Department of Electrical Engineering, University at Buffalo, Buffalo, New York, 14260, United States
*e-mail: katsumasa.yoshioka@ntt.com

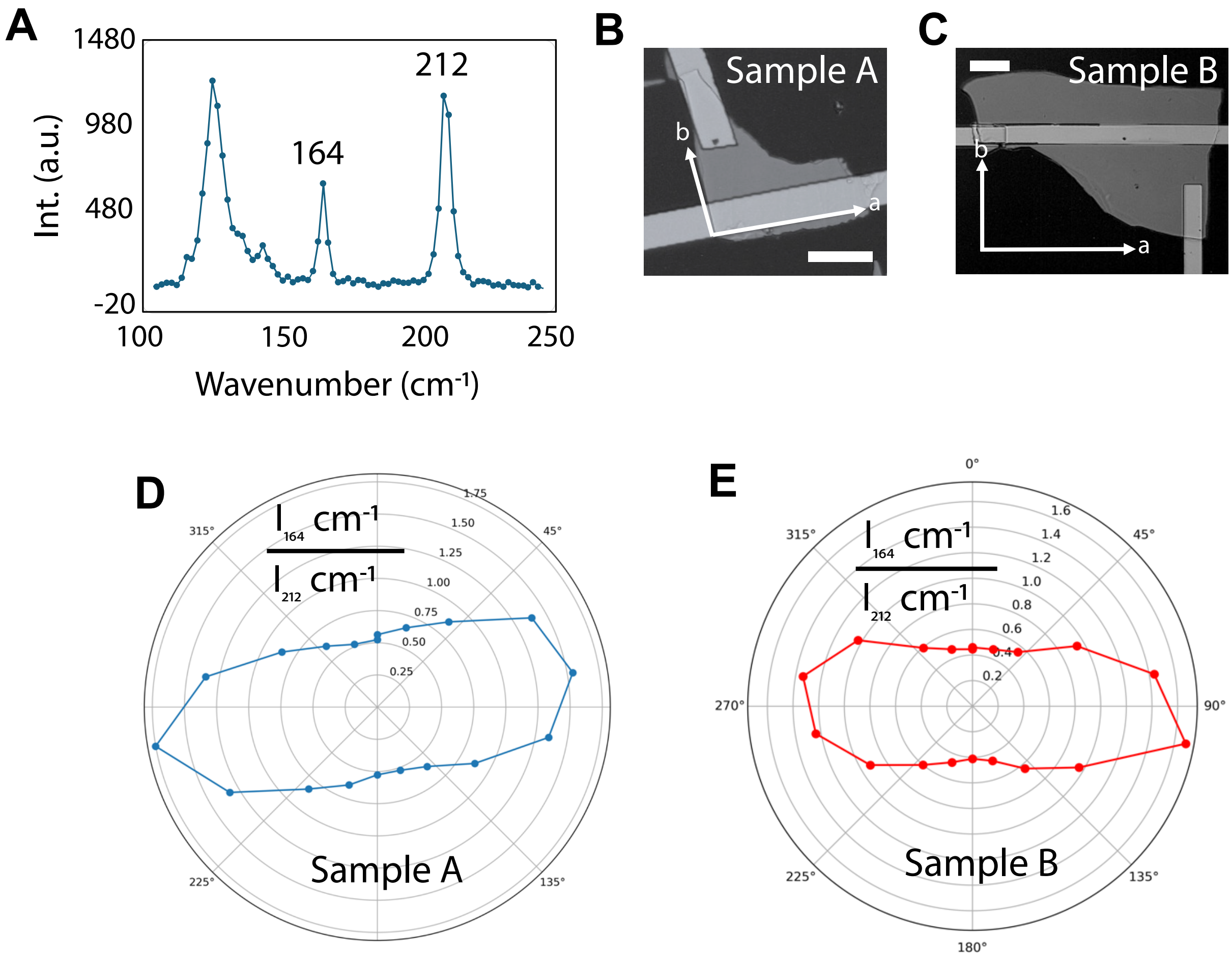

**Figure S1 | Polarization-resolved Raman spectroscopy. A,** Raman peak of $WTe_2$ flake. **B** and **C,** Optical image of the flake with axis determined by the polar plot shown in **D** and **E**, respectively. The scale bars in **B** and **C** are 20 μm.

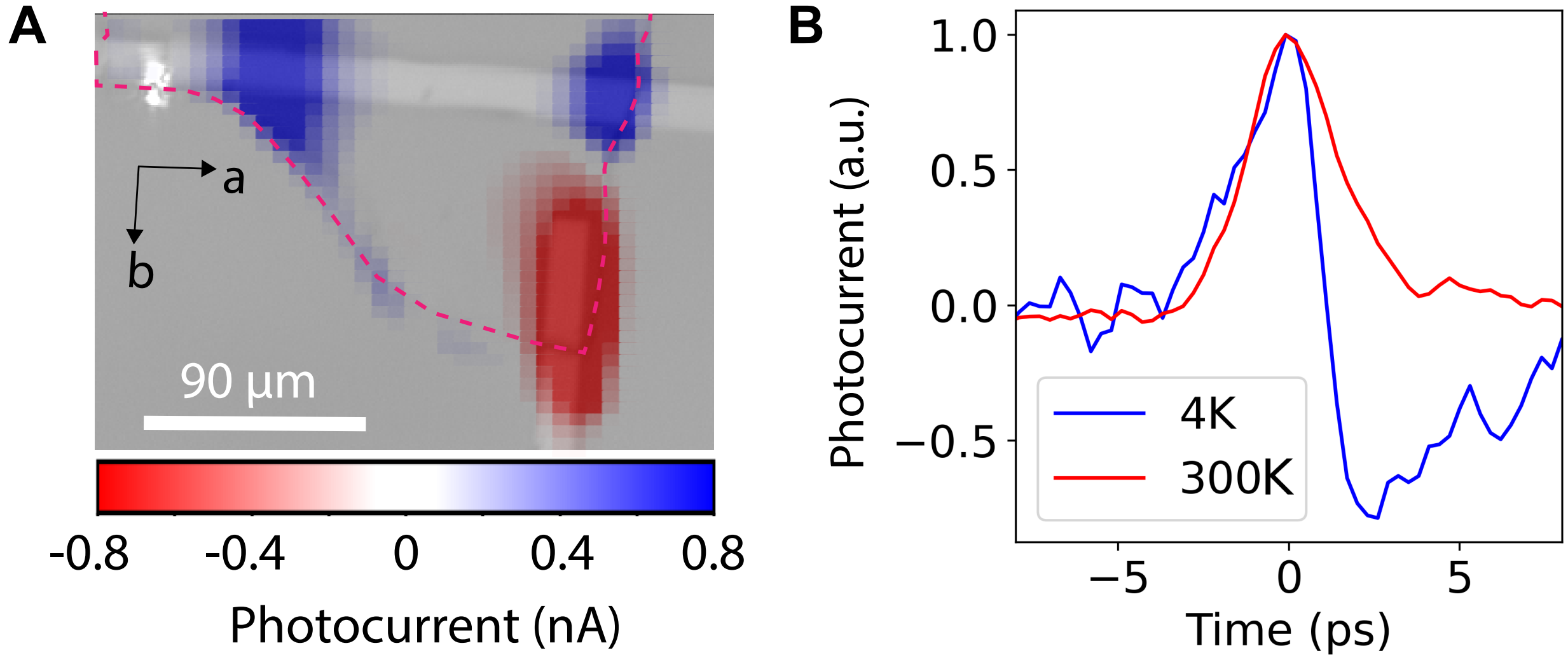

**Figure S2 | Results in Sample B. A,** Time integrated photocurrent image at 300 K. **B.** Time-domain edge photocurrent at 4 K and 300 K.

**Section I:**

**Fitting the Time domain photocurrent response:**

We follow the exponential asymmetric fitting of the time domain photocurrent from the reference.[1] The measured time domain photocurrent ($j_{meas}(t)$) was modelled as convolution of the generation ($j_{(gen)}(t)$) and the detection ($j_{(\mathrm{det})}(\mathrm{t})$) photocurrents as:

$$j_{meas}(t) \propto j_{(gen)}(t) \times j_{(det)}(-t). \qquad (1)$$

The generated current $j_{(gen)}(t)$ was expressed as the convolution of the femtosecond laser pulse $P_{pump}(t)$ with the charge carrier decay function $n_{gen}(t)$.

$$j_{gen}(t) \propto P_{(pump)}(t) \times n_{(gen)}(-t) \qquad (2)$$

$$P_{(pump)}(t) \propto e^{\frac{t^2}{2\sigma^2}} \qquad (2a)$$

$$n_{(gen)}(-t) \propto e^{\frac{t}{\tau_{rise}}} \qquad (2b)$$

In this analysis, the pulse duration of the incident femtosecond laser was taken as 280 fs, and the rise time $\tau_{rise}$ of the photoconductive switch was set to 1 ps, consistent with the response time of LT-GaAs.

The photocurrent decay time $\tau_{decay}$ attributed to the $WTe_2$ response can be extracted by fitting the measured photocurrent using an exponential asymmetric function, Equation (3). The first part of the expression corresponds to the generation process and the second part to detection:

$$j_{meas}(t) \propto e^{\frac{\sigma^2+2(t-t_0)\tau_{rise}}{2\tau_{rise}^2}} . \mathrm{erfc}\left(\frac{\sigma^2+(t-t_0)\tau_{rise}}{\sqrt{2}\sigma\tau_{rise}}\right) + e^{\frac{\sigma^2-2(t-t_0)\tau_{decay}}{2\tau_{decay}^2}} . \mathrm{erfc}\left(\frac{\sigma^2-(t-t_0)\tau_{decay}}{\sqrt{2}\sigma\tau_{decay}}\right), \quad (3)$$

where $t_0$ is the temporal position of the peak. At 4 K, the photocurrent exhibited a positive peak followed by a negative peak. To capture this behavior, the measured signal was decomposed into two components, a positive ($j_{meas}^{positive}(t)$) and a negative ($j_{meas}^{negative}(t)$), and combined as shown in Equation (4):

$$Combined\ Peak = Amplitude_{positive_peak} \times j_{meas}^{positive}(t) + Amplitude_{negative_peak} \times j_{meas}^{negative}(t). \quad (4)$$

**Section II:**

**Theoretical model for calculating bandgap, density of states, and asymmetric Seebeck coefficients:**

The first-principles calculations were performed with Quantum ESPRESSO using the generalized-gradient approximation of Perdew-Burke-Ernzerhof (PBE) for exchange and correlation. We included spin-orbit coupling and used the kinetic-energy cutoff for wavefunctions of 90 Ry. The self-consistent charge density was calculated using a k-mesh of the Brillouin zone $12 \times 6 \times 3$. In order to evaluate integrals in Equations (1) and (2) of the main text, we employed an ultra-dense mesh of $400 \times 224 \times 100$. The velocities were evaluated using fine-difference derivatives of the band structure.

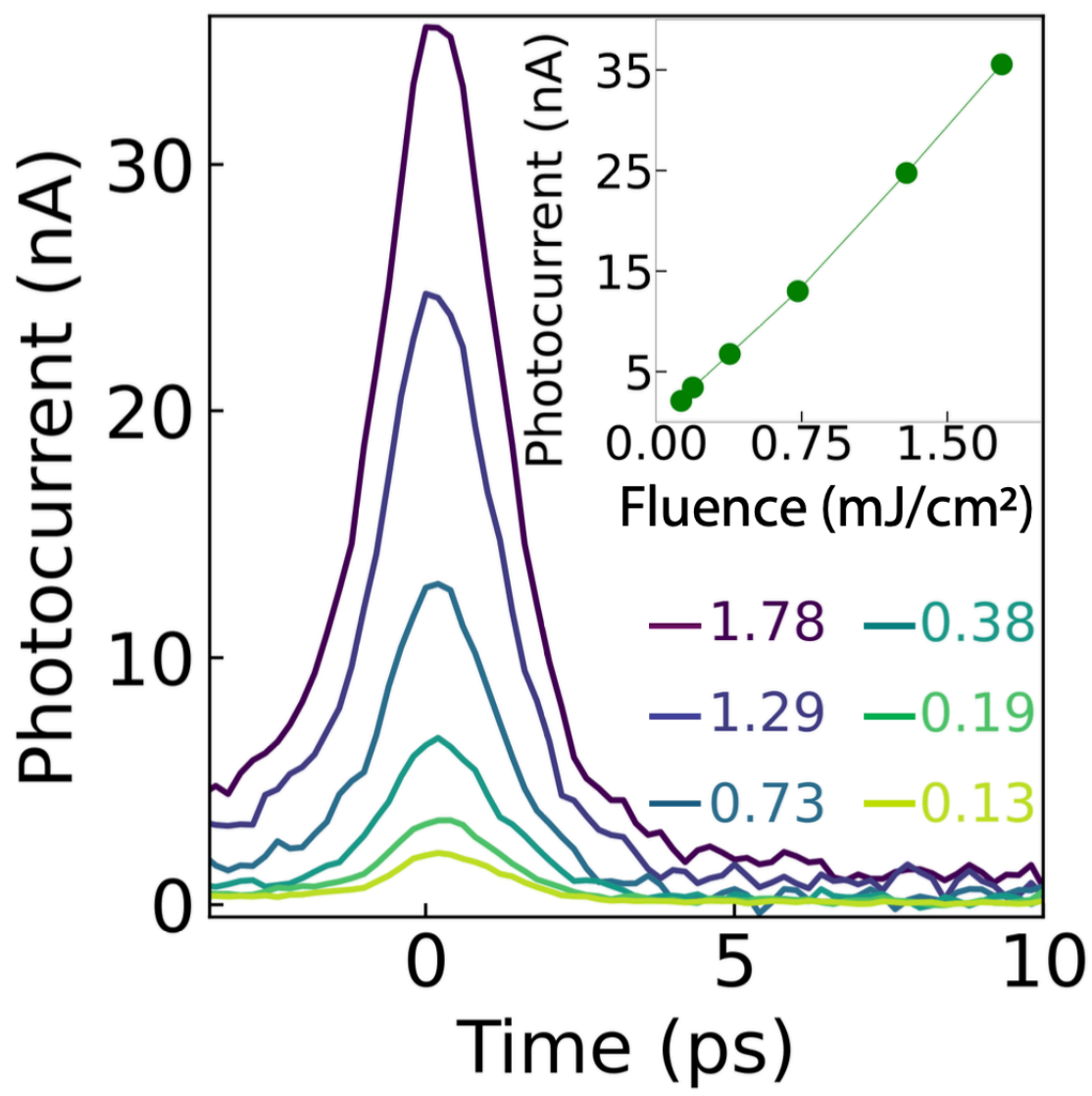

**Figure S3 | Dependence of time-domain photocurrent response on fluence at 300 K**: Time-domain response for several values of incident fluence at 300 K under linear polarization of a 517 nm femtosecond laser.

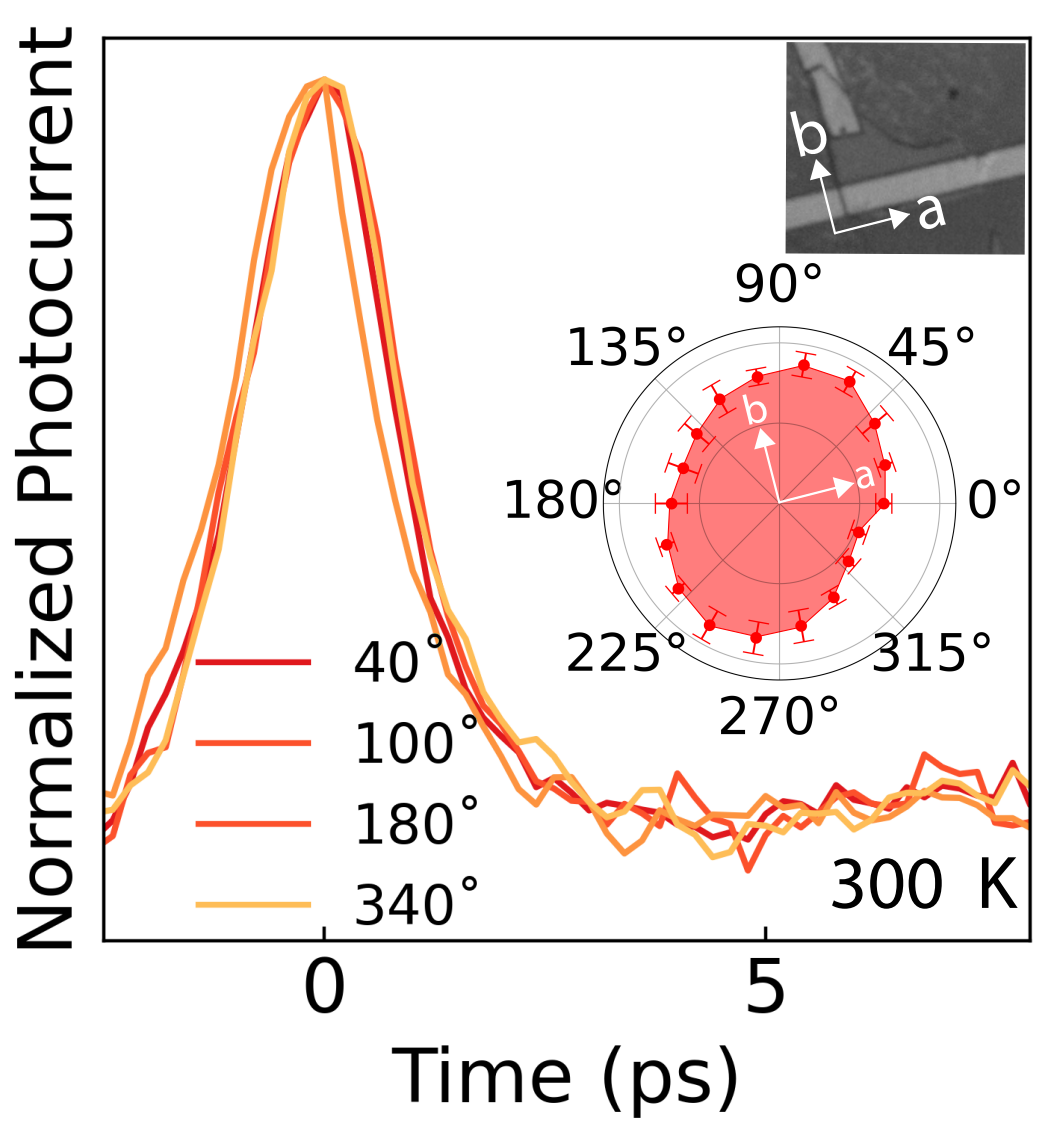

**Figure S4 | Dependence of time-domain photocurrent response on linear light polarization angle at 300 K**: Time-domain response for several linear light polarization angles at 300 K under 517 nm femtosecond laser excitation.

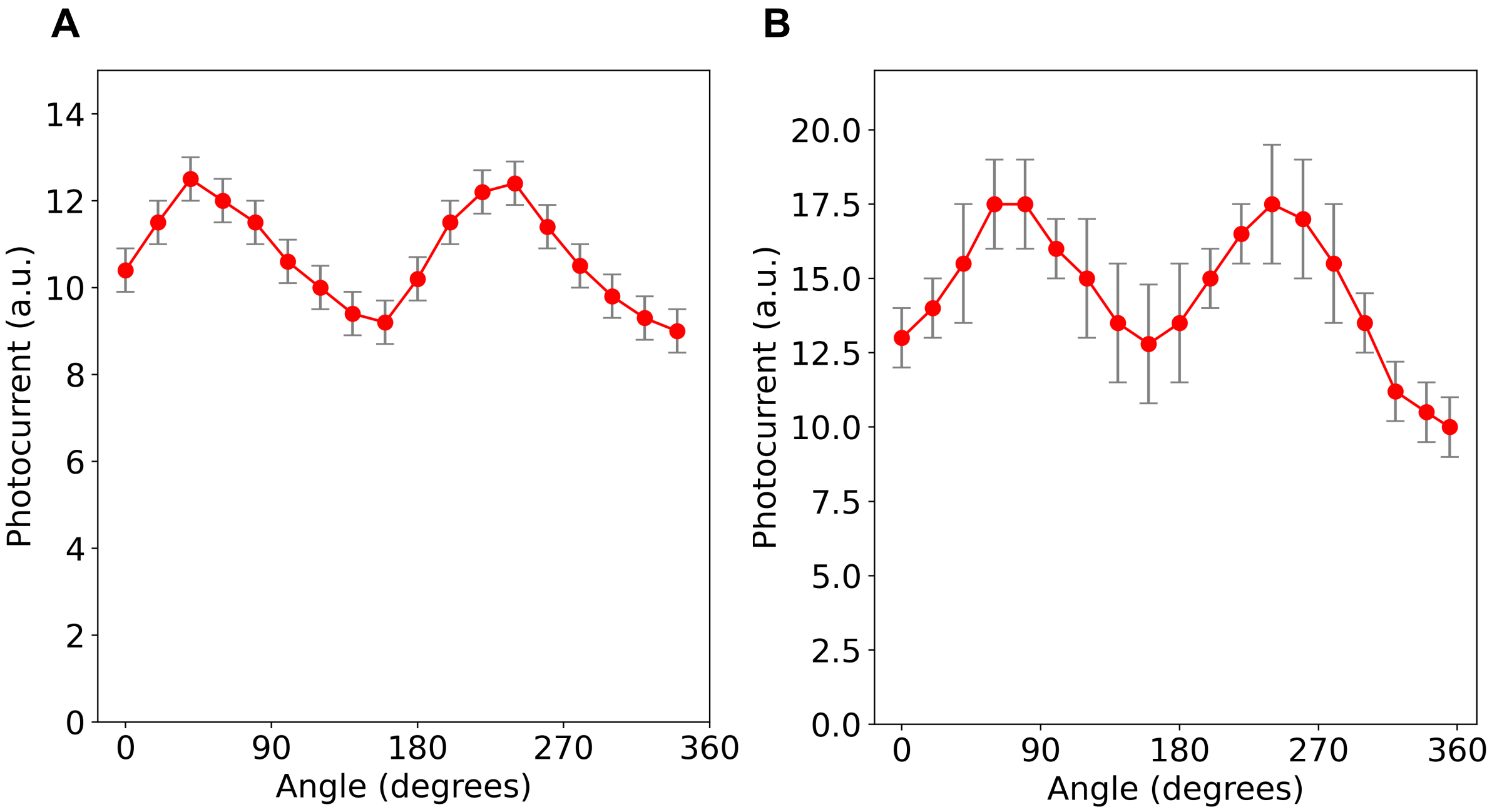

**Figure S5 | Photocurrent amplitude as a function of the polarization angle**: Photocurrent peak as a function of linear light polarization angle at 4 K (**A**) and 300 K (**B**) under 517 nm femtosecond laser excitation.

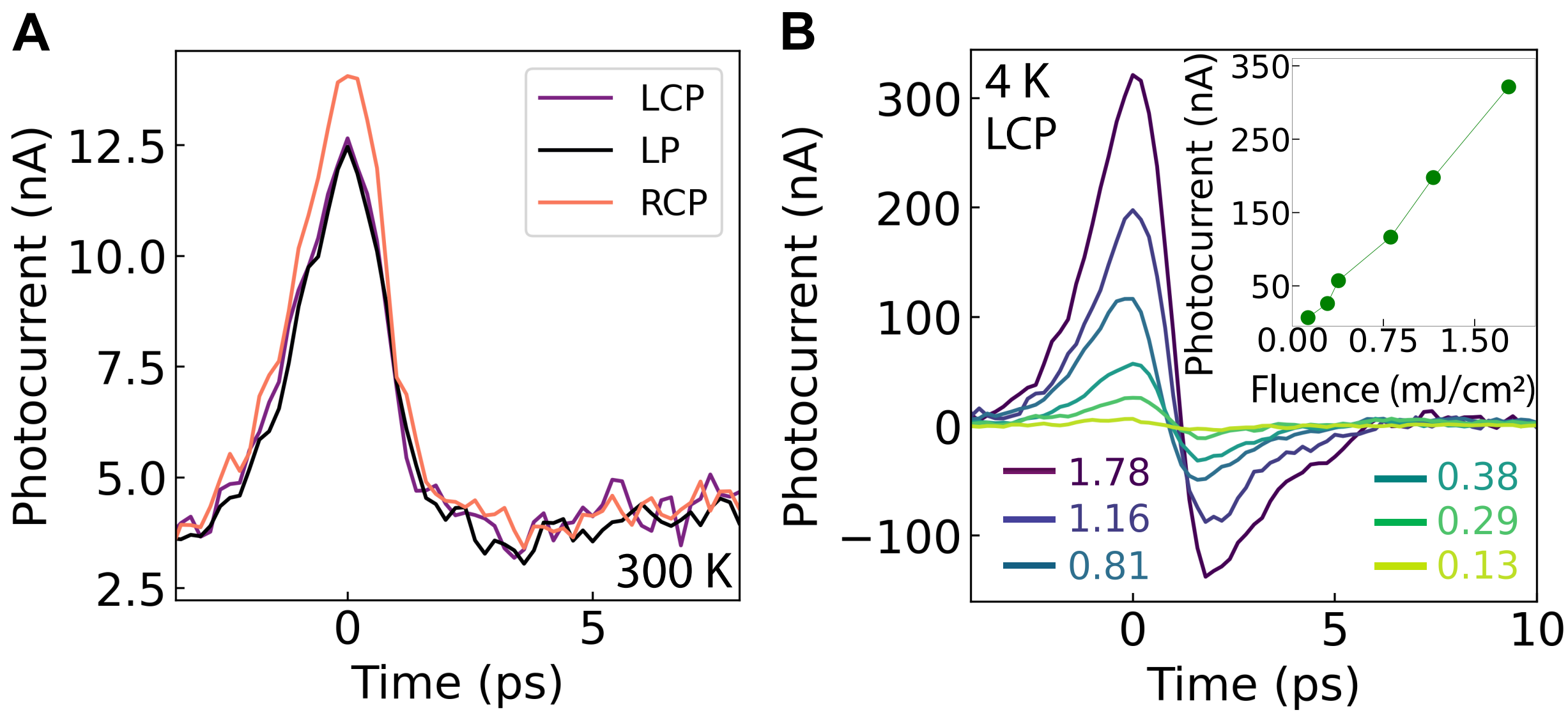

**Figure S6 | Time-domain photocurrent response under circularly polarized light**: **A,** Time-domain photocurrent response under circularly polarized light at 300 K. **B,** Time-domain photocurrent response for several values of the fluence intensity under LCP illumination at 4 K using a 517 nm femtosecond laser.